\documentclass[prl,aps,twocolumn,showpacs]{revtex4}
\usepackage{graphicx}
\usepackage{subfigure}
\begin{document}

\title{Spin-dependent Quantum Interference in \\Single-Wall Carbon Nanotubes with Ferromagnetic Contacts}
\author{H.T.~Man}
\author{I.J.W.~Wever}
\author{A.F.~Morpurgo}
\affiliation{Kavli Institute of NanoScience Delft, Faculty of
Applied Sciences, \\Delft University of Technology, Lorentzweg 1,
2628 CJ Delft, The Netherlands}
\date{\today}

\begin{abstract}
We report the experimental observation of spin-induced
magnetoresistance in single-wall carbon nanotubes contacted with
high-transparency ferromagnetic electrodes. In the linear regime the
spin-induced magnetoresistance oscillates with gate voltage in
quantitative agreement with calculations based on a
Landauer-B\"uttiker model for independent electrons. Consistent with
this interpretation, we find evidence for bias-induced oscillation
in the spin-induced magnetoresistance signal on the scale of the
level spacing in the nanotube. At higher bias, the spin-induced
magnetoresistance disappears because of a sharp decrease in the
effective spin-polarization injected from the ferromagnetic
electrodes.

\end{abstract}


\maketitle Carbon nanotubes (NTs) with ferromagnetic contacts are a
very rich and promising system for spintronics
\cite{Alphenaar_Spintronics}. Depending on the transparency of the
contacts, electronic transport through NTs occurs in different
regimes. For low transparency a NT behaves as a quantum dot and
Coulomb blockade determines the transport properties
\cite{Cobden_NT-Dot}, whereas for high transparency transport is
mainly determined by quantum interference \cite{Kong_Interference}.
This versatility gives experimental access to spin-dependent
transport phenomena that could not be investigated until now
\cite{OldNTSpin}.

Particularly important is the possibility to tune transport
electrostatically, by means of a gate voltage. The gate-voltage
dependence of spin-transport through NTs has been studied only
recently, with different experiments resulting in different
experimental observations
\cite{Jensen_SpinQD,Alphenaar_GatedSpin,Sahoo_SpinNT}. For instance,
a significant but gate-\textit{independent} influence of the contact
magnetization on the NT conductance has been observed in
Ref.~\cite{Jensen_SpinQD}. On the contrary, the experiments from
Ref.~\cite{Alphenaar_GatedSpin} show that the magnetoresistance
attributed to the spin degree of freedom is gate-voltage dependent,
but no correlation between spin-induced magnetoresistance and the
linear conductance was found. Finally, Ref.~\cite{Sahoo_SpinNT}
describes a magnetoresistance whose gate-voltage dependence
correlates with the gate-voltage dependence of the linear
conductance. For this last experiment, a full quantitative theory
based on the Landauer-B\"uttiker picture was shown to describe well
the spin- and gate-dependent magnetotransport data
\cite{Cottet_SDIPS}. Nevertheless, diverse phenomenology observed in
the different experiments calls for additional experiments, possibly
addressing different transport regimes and measurements at finite
bias, to provide an unambiguous validation of a Landauer-B\"uttiker
description of spin-transport through NTs. These experiments are
also important to convincingly exclude possible experimental
artifacts that could mimic the spin-induced magnetoresistance, such
as the magneto-Coulomb effect \cite{Shimada_MagnetoCoulomb}.

\begin{figure}[b]
  \centering
  \includegraphics[width=8.5cm]{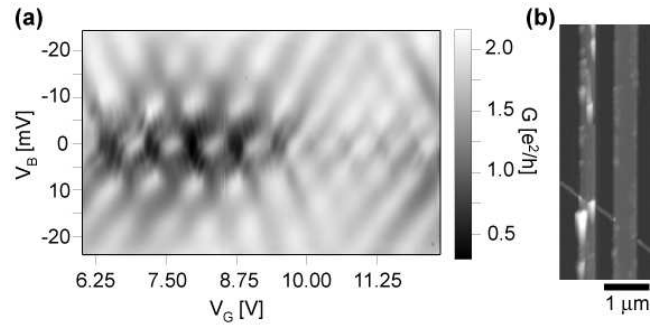}
  \caption{(a) Gate and bias dependence of the conductance measured at 4.2 K,
  exhibiting a characteristic (Fabry-Perot) pattern due to quantum interference
  of the electron waves \cite{Liang_Fabry-Perot}. (b) Atomic force microscope
  picture of the sample on which the data shown in this paper were measured.
  The obtained level spacing is approximately 6 mV, in reasonable agreement with the
  length of the nanotube between the contacts (350 nm).}\label{Figure_Cond}
\end{figure}

Here, we report experimental investigations of spin transport
through individual single-wall nanotubes with highly transparent
ferromagnetic contacts. In this regime, which is qualitatively
different from the one previously investigated (i.e.~a SWNT quantum
dot \cite{Sahoo_SpinNT}), the gate- and bias-voltage-dependent
conductance exhibits a so-called Fabry-Perot interference pattern,
originating from quantum interference of electronic waves
\cite{Liang_Fabry-Perot}. At low bias, we observe a clear
spin-induced magnetoresistance (SIMR) whose magnitude oscillates
periodically with the gate voltage. We show that the experimental
data in the linear regime compare well to predictions based on a
Landauer-B\"uttiker model for independent electrons and, at finite
energy, we find experimental evidence for the expected bias-induced
oscillations of the SIMR signal. At finite energy, we also find that
the magnitude of the SIMR signal rapidly decreases when the bias is
increased on a scale comparable to the exchange energy in the
ferromagnetic electrodes. This indicates that in actual devices the
effect of a bias-dependent spin-polarization in the contacts needs
to be taken into account.

The devices used in our investigations consist of SWNTs (either
individual or ropes) in between two PdNi contacts, prepared on a
degenerately doped silicon substrate (acting as a gate), coated with
a 500-nm-thick thermally grown oxide layer. The SWNTs are deposited
by means of a chemical vapor deposition process
\cite{Fab_Morpurgo,NTgrowth_Kong}. The electrodes consist of narrow
strips ($\sim100$ nm) of a palladium-nickel alloy electron-beam
evaporated at a base pressure of 10$^{-7}$ mbar from a mixture of
palladium and nickel in a 60\% (Pd) to 40\% (Ni) weight ratio. The
alloy is ferromagnetic above room temperature
\cite{Kontos_PdNiJunctions}. As the spin polarization of pure Nickel
is 45\%, the spin-polarization of our PdNi alloy is lower than 20\%,
and estimated to be $\simeq$ 10\%. Despite this rather low
polarization, the use of PdNi is advantageous, since the presence of
Pd enables the realization of highly transparent contacts to the NTs
\cite{PdContact_Dai, Kontos_PdNiContact}. We have investigated a
number of different samples and the data that we discuss here were
measured at 4.2 K on a sample that was sufficiently stable to
perform investigations of transport as a function of gate voltage,
bias, and magnetic field.

The dependence of the sample conductance on gate and bias voltage is
shown in Fig.~\ref{Figure_Cond}(a). The conductance oscillates in a
regular way as a function of both voltages. The resulting pattern is
similar to that of Fabry-Perot interference, which is typically
observed in ballistic SWNTs, in the presence of backscattering at
the contacts \cite{Liang_Fabry-Perot}. Spin-dependent transport in
this Fabry-Perot regime have never been addressed by previous
investigations, which have only considered the case of Coulomb
blockaded SWNTs \cite{Jensen_SpinQD,Sahoo_SpinNT}.

\begin{figure}[t]
  \centering
  \includegraphics[width=8.5cm]{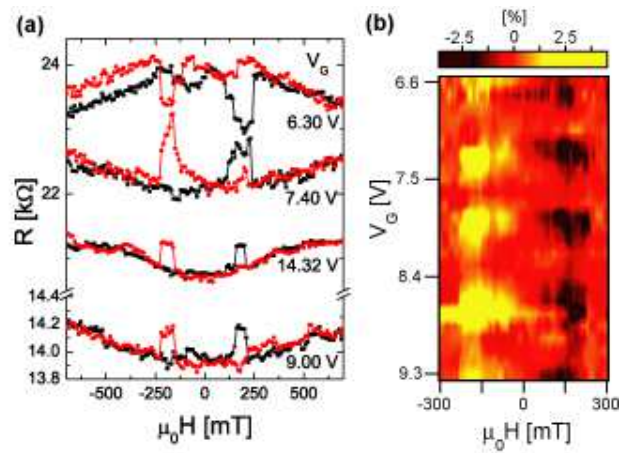}\\
  \caption{(a) Magnetoresistance traces measured at different
  gate-voltages (black dots correspond to magnetic-field up-sweeps;
  red dots to down-sweeps). The reversal of the magnetization in
  the electrodes produces an increase in magnetoresistance; in a
  few cases a decrease is observed (e.g.~$V_\mathrm{G}$= 6.30 V).
  (b) Difference between the magnetoresistance signal measured in the up-
  and down-sweeps, normalized to $R_\mathrm{P} = R$(700 mT),
  as a function of magnetic field and gate voltage. The periodicity with
  gate voltage is apparent.}\label{Figure_SIMRSel}
\end{figure}

Prior to investigating spin transport through NTs we have
characterized the switching behavior of the ferromagnetic PdNi
contacts by means of anisotropic magnetoresistance measurements. We
found that the magnetization of PdNi lies approximately in the plane
perpendicular to the strip, with a small component ($\simeq$ 10 - 15
\% of the total magnetization) pointing along the strip. Sweeping
the magnetic field in the plane perpendicular to the strip results
in a magnetization reversal for $|\mu_0 H|$ in between 100 and 300
mT, with the exact value depending on the width of the strip. At low
field, a small, continuous change of the magnetization orientation
also occurs due to the change of the in-plane component, since the
complete alignment of the magnetization along an applied magnetic
field requires a field of several tesla. For the SIMR measurements
on SWNTs this behavior of the magnetization should result in
hysteretic features at the magnetic field values for which the
magnetization reversal occurs, superimposed onto smooth changes in
the resistance due to the gradual change in magnetization
orientation in the two PdNi leads \cite{note1}.

Figure \ref{Figure_SIMRSel}(a) shows magnetoresistance curves for
the SWNT sample measured for different values of the gate voltage,
for an applied magnetic field increasing from $-700$ to 700 mT
(up-sweep) and subsequently decreasing from 700 to $-700$ mT
(down-sweep). The width of the two PdNi contacts in this sample are
150 nm and 500 nm, resulting in magnetization reversal at magnetic
field values of 250 and 125 mT respectively. A hysteretic feature in
the resistance is seen in the expected magnetic field range,
corresponding to an antiparallel orientation of the magnetization in
the PdNi electrodes. The feature is superimposed on a smoother
background, as expected. The absolute change in resistance has a
magnitude of a few tenths of a kilo-ohm, which is much larger than
the total resistance of the PdNi strips. This implies that the
effect cannot be accounted for in terms of a change in the
resistance of part of the PdNi contacts, which is much lower. The
change in magnetoresistance induced by the magnetization reversal is
positive for most values of gate voltage, i.e. the antiparallel
orientation of the magnetization in the contacts results in an
increase of the device resistance. In a few cases, however, a
negative magnetoresistance has been observed
\cite{Sahoo_SpinNT,Cottet_SDIPS} (see Fig.~\ref{Figure_SIMRSel}(a),
top curve).

To investigate the SIMR in more detail, we have measured up- and
downsweep linear magnetoresistance traces for $V_\mathrm{G}$ between
6.6 and 9.3 V, corresponding to approximately four periods of the
gate-voltage-induced conductance oscillations.
Fig.~\ref{Figure_SIMRSel}(b) shows the difference between the
resistance measured in the up- and down-magnetic-field sweeps as a
function of $V_\mathrm{G}$ and $\mu_0 H$, normalized to the
resistance at $\mu_0 H =$ 700 mT. It is apparent that the magnitude
of the SIMR oscillates with gate voltage \textit{in phase} with the
conductance, which excludes the possibility that the observed
magnetoresistance is due to a magneto-coulomb effect
\cite{Shimada_MagnetoCoulomb}. The period of the oscillations also
corresponds to that of the total conductance, thus demonstrating
that the SIMR originates from quantum interference of electron waves
backscattered at the contacts.

\begin{figure}[t]
  \centering
  \includegraphics[width=8cm]{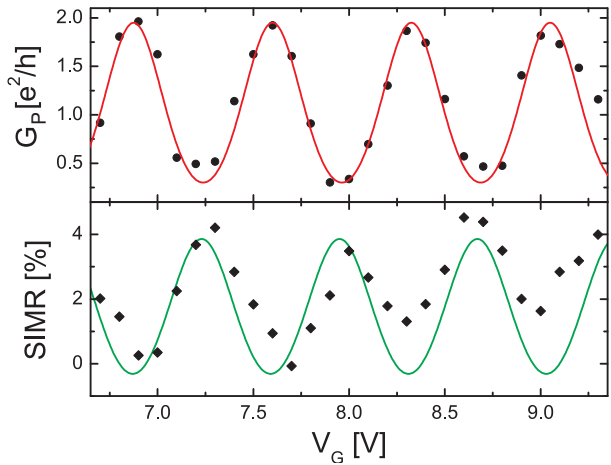}\\
  \caption{Comparison of theoretical predictions based on Eqs.~\ref{Eq_TotalCond} and \ref{Eq_Trans} (lines) and
  experimental data (symbols) for the gate-voltage dependent
  conductance (top) and amplitude of the spin-induced magnetoresistance (bottom).}
\label{Figure_FitCondAndMR}
\end{figure}

From the magnetoresistance measurements we also extract the
conductance $G_\mathrm{P} = G(V_\mathrm{G}, 700$ mT) and the
relative magnitude of the SIMR
$\frac{R_\mathrm{AP}(V_\mathrm{G})-R_\mathrm{P}(V_\mathrm{G})}{R_\mathrm{P}(V_\mathrm{G})}$
(see Fig.~\ref{Figure_FitCondAndMR}). For
$R_\mathrm{AP}(V_\mathrm{G})$ we take the value of the resistance
measured at $\mu_0 H$ = 175 mT, i.e.~in the middle of the hysteretic
feature originating from the magnetization reversal. For
$R_\mathrm{P}(V_\mathrm{G})$ we take the value measured at $\mu_0 H
= 700$ mT.

To interpret the linear magnetoresistance data, we adopt a
Landauer-B\"uttiker picture. The spin-dependent, zero-temperature
conductance of an individual SWNT can then be written as a function
of the transmission probability for the majority spin ($+$) and
minority spin ($-$) orientations of the left ($s$) and right ($r$)
contact as
\begin{equation}\label{Eq_TotalCond}
G = \frac{2 e^2}{h} \sum_{s,r\in\{+,-\}} T_{sr}
\end{equation}

\noindent (the factor of 2 accounts for the doubly degenerate bands
of SWNTs). Since in SWNTs spin-orbit interaction is negligibly
small, we only have to consider transmission from $(+)_s$ to $(+)_r$
and $(-)_s$ to $(-)_r$ when the magnetization in the contacts point
parallel to each other (i.e.~$G_\mathrm{P} = 2 e^2/h
(T_{++}+T_{--})$), and $(+)_s \rightarrow (-)_r$ and $(-)_s
\rightarrow (+)_r$ spin (i.e.~$G_\mathrm{AP} = 2 e^2/h
(T_{+-}+T_{-+})$) when the magnetization vectors are anti-parallel.

For ballistic propagation in the SWNT, the transmission probabilities read:

\begin{equation}\label{Eq_Trans}
T_{sr} = \frac{T_L^s T_R^r}{|1 - [(1 - T_L^s)(1 - T_R^r)]^{1/2}
e^{2i\delta}|^2},
\end{equation}

\noindent where, following Refs.~\cite{Cottet_SDIPS,Sahoo_SpinNT},
we write the spin-dependent transmission probabilities $T^s_{L(R)}$
at the left and right contact as $T^s_{L(R)} = T_{L(R)}(1 + s P)$,
with $s = \pm 1$ and $P$ the magnitude of the polarization in the
contacts ($P= 0.1$ as it is appropriate for our PdNi electrodes).
The quantity $\delta$ is the phase acquired by one electron during
its propagation from one contact to the other and back. It includes
the spin-dependent phase acquired during the scattering process
\cite{Cottet_SDIPS}, and the dynamical phase, which is linearly
dependent on gate voltage and electron energy. We set the phase
acquired in the reflections at the contacts to zero, to decrease the
number of fitting parameters.

\begin{figure}[b]
  \centering
  \includegraphics[width=8.5cm]{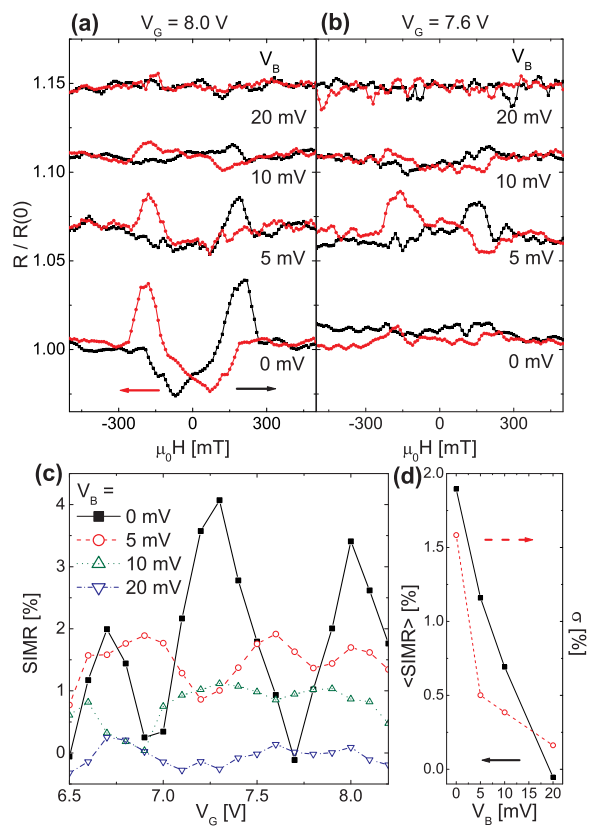}\\
  \caption{Panels (a) and (b): magnetoresistance traces for different
  values of the applied bias $V_\mathrm{B}$, at $V_\mathrm{G}$ = 8.0 and
  $V_\mathrm{G}$ = 7.6 V (curves offset for clarity). At $V_\mathrm{G}$ =
  8.0 V the SIMR signal at  $V_\mathrm{B}$ = 0 mV is larger than at
  $V_\mathrm{B}$ = 5 mV; at $V_\mathrm{G}$ = 7.6 V the situation is reversed.
  This indicates that the amplitude of the SIMR signal oscillates with bias.
  The full $V_\mathrm{G}$-dependence is shown in (c). Panels (a-c) also clearly
  show that the SIMR signal at $V_\mathrm{B}$ = 20 mV is fully suppressed.
  Panel (d): Bias-dependence of the gate-voltage averaged SIMR and of its standard
  deviation (as a measure of the oscillating component). }\label{Figure_TotalBias}
\end{figure}

By convoluting Eq.~\ref{Eq_TotalCond} with the Fermi distribution,
we evaluate the finite-temperature conductance and SIMR signal,
using $T^s_{L(R)}$  as the only free parameters. The solid lines in
Fig.~\ref{Figure_FitCondAndMR} correspond to the theoretical
predictions obtained with $T_L$ = 0.84 and $T_R$ = 0.26, i.e.~the
contacts exhibit only a small asymmetry. It is apparent that the
model reproduces correctly the observed behavior of the linear
conductance, the absolute magnitude of the SIMR signal ($\sim$4 \%),
and its periodicity for the entire measured gate-voltage range with
only these two parameters. The small quantitative
deviations seen in the detailed gate-voltage dependence of the SIMR
signal can be explained (at least in part) by the misorientation of
the magnetization in the two electrodes.

To understand up to what extent a Landauer-B\"uttiker picture
describes spin-dependent transport in SWNTs, we have also performed
SIMR measurements at finite energy. The magnitude of the SIMR effect
measured in the differential conductance should oscillate as a
function of bias for independent electrons in a SWNT in the
Fabry-Perot regime, since the electron energy enters the phase
$\delta$ linearly, similarly to the gate voltage.
Fig.~\ref{Figure_TotalBias}(a) and \ref{Figure_TotalBias}(b) show
magnetoresistance traces measured at $V_\mathrm{B}$ = 0 and 5 mV,
for two values of the gate voltage. Whereas at $V_\mathrm{G}$ = 8.0
V the SIMR signal at $V_\mathrm{B}$ = 0 is larger than that at
$V_\mathrm{B}$ = 5 mV, at $V_\mathrm{G}$ = 7.6 V the opposite is
observed. This shows that oscillations as a function of bias on the
scale of the level spacing in the nanotube (see Fig. 1) are indeed
present, as further illustrated in Fig.~\ref{Figure_TotalBias}(c)
(curves at 0 and 5 mV), where the magnitude of the SIMR signal is
plotted versus $V_\mathrm{G}$. This observation, in conjunction with
the fact that also at finite bias the SIMR oscillates in phase with
the gate-dependent differential conductance as expected from
Eqs.~\ref{Eq_TotalCond} and \ref{Eq_Trans}, indicate that a
Landauer-B\"uttiker picture provide a good first approximation to
describe the energy dependence of spin-dependent transport in SWNTs.

Fig.~\ref{Figure_TotalBias}, however, also clearly shows that the
magnitude of the SIMR signal rapidly decreases with increasing
voltage. As a consequence, at $V_\mathrm{B}$= 20 mV no SIMR is
detected. Interestingly (see Fig.~\ref{Figure_TotalBias}(d)), the
rapid suppression with bias occurs for both the
$V_\mathrm{G}$-dependent oscillations, originating from quantum
interference, as well as for the gate-voltage averaged signal, which
corresponds to the "classical" SIMR signal. These observations
deviate from the behavior expected on the basis of
Eqs.~\ref{Eq_TotalCond} and \ref{Eq_Trans}, which predicts that the
$V_\mathrm{G}$-induced oscillations in the SIMR signal should
preserve their amplitude at high bias. Observing that the
"classical" (i.e., gate-averaged) SIMR signal disappears at high
energy suggests that the suppression of SIMR with bias does not
originate from a break-down of the Landauer-B\"uttiker picture, but
rather from a decrease of the spin-polarization in the electrodes,
i.e. the polarization $P$ decreases with increasing bias. This
interpretation is suggested by the fact that the bias at which the
SIMR signal disappears ($\approx 20$ meV) is comparable to the
exchange energy in the PdNi electrodes used in the experiments
($E_\mathrm{ex} \simeq 30$ mV) \cite{Kontos_PdNiJunctions}. This is
consistent with the known behavior of the magnetoresistance of
tunnel junctions fabricated with many other weakly ferromagnetic
alloys (e.g., Cu-Ni), which also show a rapid decrease of the
effective spin polarization with increasing the bias on the scale of
the exchange energy\cite{CuNiBiasDep_Marrows}.

In conclusion, our systematic investigation of spin-dependent
transport as a function of bias and gate voltage indicate that the
observed phenomenology can be described in terms of a
Landauer-B\"uttiker picture, provided that the bias dependence of
the spin polarization injected from the ferromagnetic contacts is
taken into account.

We acknowledge fruitful discussions with A.~Cottet,
S.T.B.~G\"onnenwein, T.M.~Klapwijk, T.~Kontos, C.~Strunk, and
F.~Zwanenburg. We thank S.~Sapmaz for experimental help and
C.~Dekker for the use of laboratory equipment. We are grateful to
FOM and NWO (Vernieuwingsimpuls 2000 program) for the financial
support.


\begin{thebibliography}{99}
\bibitem{note1} We refer to the magnetization orientation as
"parallel" and "anti-parallel", despite the presence of the small
in-plane component.
\end{thebibliography}
\end{document}